\documentstyle[12pt]{article}
\begin{document}
\def\beq{\begin{equation}}
\def\eeq{\end{equation}}
\def\bea{\begin{eqnarray}}
\def\eea{\end{eqnarray}}
\def\ve{\vert}
\def\vel{\left|}
\def\ver{\right|}
\def\nnb{\nonumber}
\def\ga{\left(}
\def\dr{\right)}
\def\aga{\left\{}
\def\adr{\right\}}
\def\rar{\rightarrow}
\def\nnb{\nonumber}
\def\la{\langle}
\def\ra{\rangle}
\def\lla{\left<}
\def\rra{\right>}
\def\ba{\begin{array}}
\def\ea{\end{array}}
\def\tep{$B \rar K \ell^+ \ell^-$}
\def\tepm{$B \rar K \mu^+ \mu^-$}
\def\tept{$B \rar K \tau^+ \tau^-$}
\def\ds{\displaystyle}


\def\simlt{\stackrel{<}{{}_\sim}}
\def\simgt{\stackrel{>}{{}_\sim}}


\title{ {\Large {\bf 
Rare $B \rar \ell^+ \ell^- \gamma$ decay and new physics effects 
        } 
                }       
      }

\author{\vspace{1cm}\\
{\small T. M. Aliev \thanks
{e-mail: taliev@metu.edu.tr}\,\,,
A. \"{O}zpineci \thanks
{e-mail: altugoz@metu.edu.tr}\,\,,
M. Savc{\i} \thanks
{e-mail: savci@metu.edu.tr}} \\
{\small Physics Department, Middle East Technical University} \\
{\small 06531 Ankara, Turkey} }
\date{}

\begin{titlepage}
\maketitle
\thispagestyle{empty}

\begin{abstract}
Using the most general, model independent form of the effective Hamiltonian
rare decays $B \rar \ell^+ \ell^- \gamma~(\ell=\mu,~\tau)$ are studied.
The sensitivity of the photon energy distribution and branching ratio to the
new Wilson coefficients is investigated.
\end{abstract}

~~~PACS number(s): 12.60.--i, 13.20.--v, 13.20.He
\end{titlepage}

\section{Introduction}
Started to work, the two $B$--factories open an excited new era in studying
$B$ meson decays \cite{R1,R2}. The main research program of these factories
is studying CP violation in $B$ meson system and investigating their decays.
From theoretical point of view, interest to the rare decays can be
attributed to the fact that they occur at loop level in the Standard Model 
(SM) and they are very sensitive to the flavor structure of the SM as well 
as to the new physics beyond the SM. From experimental point of view
studying radiative $B$ meson decays can provide us essential information
on the parameters of the SM, such as the elements of the 
Cabibbo--Kobayashi--Maskawa (CKM) matrix, the leptonic decay constants 
etc., which are yet poorly known.

It is well known that the flavor--changing neutral current process
$B_{s(d)} \rar \ell^+ \ell^-$ has helicity suppression. These decays are
proportional to the lepton mass and because of this reason the
decay width of these processes are too small to be measured for the light
lepton modes. It should be noted that in the SM the branching ratio of the
${\cal B}(B_s \rar e^+ e^-,~\mu^+ \mu^-)\simeq 4.2 \times 10^{-14}$ and
$1.8 \times 10^{-9}$, respectively. Although $\tau$ channel is free of this
suppression, its experimental detection is quite hard due to the low
efficiency. It has been observed that the radiative leptonic
$B^+ \rar \ell^+ \nu \gamma~(\ell=e,~\mu)$ decays have larger branching
ratio compared to that of the purely leptonic models \cite{R3}--\cite{R9}.
It was shown in \cite{R10,R11} that similar situation takes place for the
radiative decays $B_{s(d)} \rar \ell^+ \ell^- \gamma$. In these decays the
contribution of the diagram when photon is radiated from an intermediate 
charged line,
can be neglected, since it is strongly suppressed by a factor $m_b^2/m_W^2$.
Moreover, the internal Bremsstrahlung (IB) part when photon is emitted from
external charged leptons is proportional to lepton mass, which follows from
helicity arguments, gives small contribution. For this reason in 
$B \rar \ell^+ \ell^- \gamma$ decay the main contribution should come from
the diagrams, when photon is emitted from the initial quarks, i.e.,
structure--dependent part (SD), since they are free of the helicity
suppression. Therefore the decay rate of the $B_{s(d)} \rar \ell^+ \ell^-
\gamma~(\ell=e,~\mu)$ might have an enhancement in comparison to the pure   
leptonic models of $B_{s(d)} \rar \ell^+ \ell^-$ decay if the SD
contributions to the decays are dominant and hence 
$B_q \rar \ell^+ \ell^- \gamma$ decay might be sensitive to
the new physics effects beyond SM. New physics effects in rare $B_q$ decays
can appear in two different ways; either through new contributions to the
Wilson coefficients existing in the SM or through the new operators in the
effective Hamiltonian which are absent in the SM. The goal of this work is
combining both these approaches to study the sensitivity of of the
physically measurable quantities, like branching ratio, photon energy
distribution, to the new physics effects.

The work is organized as follows. In section 2, we derive the general
expression for the photon energy distribution using the most general form of
four--Fermi interaction. In section 3 we investigate the sensitivity of
photon energy distribution and branching ratio to the new Wilson
coefficients. 

\section{Matrix element for the $B_q \rar \ell^+ \ell^- \gamma$ decay}
In this section we calculate the photon energy distribution and branching
ratio for the $B_q \rar \ell^+ \ell^- \gamma$ decay using the most general
model independent form of the effective Hamiltonian. The matrix element for
the process $B \rar \ell^+ \ell^- \gamma$ can be obtained from that of the
purely leptonic $B \rar \ell^+ \ell^-$ decay. The effective $b\rar q \ell^+
\ell^-$ transition can be written in terms of twelve model independent
four--Fermi interactions can be written in the following form \cite{R12}:
\bea
\label{effH}
{\cal H}_{eff} &=& \frac{G\alpha}{\sqrt{2} \pi}
 V_{tq}V_{tb}^\ast
\Bigg\{ C_{SL} \, \bar q i \sigma_{\mu\nu} \frac{q^\nu}{q^2}\, L \,b
\, \bar \ell \gamma^\mu \ell + C_{BR}\, \bar q i \sigma_{\mu\nu}
\frac{q^\nu}{q^2} \,R\, b \, \bar \ell \gamma^\mu \ell \nnb \\
&&+C_{LL}^{tot}\, \bar q_L \gamma_\mu b_L \,\bar \ell_L \gamma^\mu \ell_L +
C_{LR}^{tot} \,\bar q_L \gamma_\mu b_L \, \bar \ell_R \gamma^\mu \ell_R +
C_{RL} \,\bar q_R \gamma_\mu b_R \,\bar \ell_L \gamma^\mu \ell_L \nnb \\
&&+C_{RR} \,\bar q_R \gamma_\mu b_R \, \bar \ell_R \gamma^\mu \ell_R +
C_{LRLR} \, \bar q_L b_R \,\bar \ell_L \ell_R +
C_{RLLR} \,\bar q_R b_L \,\bar \ell_L \ell_R \\
&&+C_{LRRL} \,\bar q_L b_R \,\bar \ell_R \ell_L +
C_{RLRL} \,\bar q_R b_L \,\bar \ell_R \ell_L+
C_T\, \bar q \sigma_{\mu\nu} b \,\bar \ell \sigma^{\mu\nu}\ell \nnb \\
&&+i C_{TE}\,\epsilon^{\mu\nu\alpha\beta} \bar q \sigma_{\mu\nu} b \,
\bar \ell \sigma_{\alpha\beta} \ell  \Bigg\}~, \nnb
\eea
where the chiral projection operators $L$ and $R$ in (\ref{effH}) are
defined as
\bea
L = \frac{1-\gamma_5}{2} ~,~~~~~~ R = \frac{1+\gamma_5}{2}\nnb~,
\eea
and $C_X$ are the coefficients of the four--Fermi interactions.
It can easily be seen from Eq. (\ref{effH}) that
several of all Wilson coefficients do already exist in the SM. The
coefficients $C_{SL}$ and $C_{BR}$ correspond to $-2 m_s C_7^{eff}$ and $-2
m_b C_7^{eff}$ in the SM, respectively. The next four terms in this
expression are the vector interactions. The interaction terms containing
$C_{LL}^{tot}$ and $C_{LR}^{tot}$ exist in the SM in the form
$C_9^{eff}-C_{10}$ and $C_9^{eff}+C_{10}$, respectively. Therefore
$C_{LL}^{tot}$ and $C_{LR}^{tot}$ describe the contributions coming from the
SM and the new physics, whose explicit are
\bea
C_{LL}^{tot} &=& C_9^{eff} - C_{10} + C_{LL}~, \nnb \\
C_{LR}^{tot} &=& C_9^{eff} + C_{10} + C_{LR}~. \nnb
\eea
The terms with coefficients $C_{LRLR}$, $C_{RLLR}$, $C_{LRRL}$ and 
$C_{RLRL}$ describe the scalar type interactions. The last two 
terms in Eq. (\ref{effH}) with the coefficients $C_T$ and $C_{TE}$
describe the tensor type interactions.

Having presented the general form of the effective Hamiltonian the next
problem is calculation of the matrix element of the $B_q \rar \ell^+ \ell^-
\gamma$ decay. This matrix element can be written as the sum of the
structure--dependent and internal Bremsstrahlung parts 
\bea
{\cal M} = {\cal M}_{SD}+{\cal M}_{IB}~. 
\eea
It follows from Eq. (\ref{effH}) that, in order to calculate the matrix 
element ${\cal M}_{SD}$ for the structure--dependent part, the following 
matrix elements are needed
\bea
\label{mel}
&&\lla \gamma\vel \bar s \gamma_\mu (1 \mp \gamma_5)
b \ver B \rra~,\nnb \\
&&\lla \gamma \vel \bar s \sigma_{\mu\nu} b \ver B \rra~, \nnb \\
&&\lla \gamma \vel \bar s (1 \mp \gamma_5) b
\ver B \rra~.
\eea
The first two of the matrix elements in Eq. (\ref{mel}) are defined as
\cite{R4,R10}
\bea
\label{mel1}
\lla \gamma(k) \vel \bar q \gamma_\mu
(1 \mp \gamma_5) b \ver B(p_B) \rra &=&
\frac{e}{m_B^2} \Big\{
\epsilon_{\mu\nu\lambda\sigma} \varepsilon^{\ast\nu} q^\lambda
k^\sigma g(q^2) \nnb \\
&&\pm i\,
\Big[ \varepsilon^{\ast\mu} (k q) -
(\varepsilon^\ast q) k^\mu \Big] f(q^2) \Big\}~,\\ \nnb \\
\label{mel2}
\lla \gamma(k) \vel \bar q \sigma_{\mu\nu} b \ver B(p_B) \rra &=& 
\frac{e}{m_B^2}
\epsilon_{\mu\nu\lambda\sigma} \Big[
G \varepsilon^{\ast\lambda} k^\sigma +
H \varepsilon^{\ast\lambda} q^\sigma +
N (\varepsilon^\ast q) q^\lambda k^\sigma \Big]~,
\eea
respectively,
where $\varepsilon_\mu^\ast$ and $k_\mu$ are the four vector polarization
and four momentum of the photon, respectively, $q$ is the momentum transfer
and $p_B$ is the momentum of the $B$ meson.
The matrix element 
$\lla \gamma(k) \vel \bar s \sigma_{\mu\nu} \gamma_5 b \ver B(p_B) \rra$
can be obtained from Eq. (\ref{mel2}) using the identity
\bea
\sigma_{\mu\nu} = - \frac{i}{2}\epsilon_{\mu\nu\alpha\beta}
\sigma^{\alpha\beta} \gamma_5~.\nnb
\eea
The matrix elements 
$\lla \gamma(k) \vel \bar s (1 \mp \gamma_5) b \ver B(p_B) \rra$ 
and
$\lla \gamma \vel \bar s i \sigma_{\mu\nu} q^\nu b \ver B \rra$ can be
calculated by contracting both sides of the  Eqs. (\ref{mel1}) and (\ref{mel2})
with $q_\nu$, respectively. We get then
\bea
\label{mel1q}
\lla \gamma(k) \vel \bar s (1 \mp \gamma_5) b \ver B(p_B) \rra
&=& 0~, \\
\label{mel2q}
\lla \gamma \vel \bar s i \sigma_{\mu\nu} q^\nu b \ver B \rra &=&
\frac{e}{m_B^2} i\, \epsilon_{\mu\nu\alpha\beta} q^\nu
\varepsilon^{\alpha\ast} k^\beta G~.
\eea
Using Eqs. (\ref{mel2})and (\ref{mel2q}) the matrix element $\lla \gamma
\vel \bar s i \sigma_{\mu\nu} q^\nu (1+\gamma_5) b \ver B \rra$ can be
written in terms of form factors that are calculated in framework of the QCD
sum rules \cite{R10} as follows 
\bea
\label{mel22q}
\lla \gamma \vel \bar s i \sigma_{\mu\nu} q^\nu (1+\gamma_5) b \ver B \rra &=&
\frac{e}{m_B^2} \Big\{
\epsilon_{\mu\alpha\beta\sigma} \, \varepsilon^{\alpha\ast} q^\beta k^\sigma
g_1(q^2)
+ i\,\Big[\varepsilon_\mu^\ast (q k) - (\varepsilon^\ast q) k_\mu \Big]
f_1(q^2) \Big\}~.
\eea
It should be noted that these form factors were calculated in framework of
the light--front model in \cite{R13}.
So, using Eqs. (\ref{mel2}), (\ref{mel2q}) and (\ref{mel22q}) we can easily
express $G,~H$ and $N$ in terms of the form factors $g_1$ and $f_1$. The
matrix element which describes the structure--dependent part can be obtained 
from Eqs. (\ref{mel1})--(\ref{mel22q})
\bea
\label{sd}
{\cal M}_{SD} &=& \frac{\alpha G_F}{4 \sqrt{2} \, \pi} V_{tb} V_{tq}^* 
\frac{e}{m_B^2} \,\Bigg\{
\bar \ell \gamma^\mu (1-\gamma_5) \ell \, \Big[
A_1 \epsilon_{\mu \nu \alpha \beta} 
\varepsilon^{\ast\nu} q^\alpha k^\beta + 
i \, A_2 \Big( \varepsilon_\mu^\ast (k q) - 
(\varepsilon^\ast q ) k_\mu \Big) \Big] \nnb \\
&+& \bar \ell \gamma^\mu (1+\gamma_5) \ell \, \Big[
B_1 \epsilon_{\mu \nu \alpha \beta} 
\varepsilon^{\ast\nu} q^\alpha k^\beta 
+ i \, B_2 \Big( \varepsilon_\mu^\ast (k q) - 
(\varepsilon^\ast q ) k_\mu \Big) \Big] \nnb \\
&+& i \, \epsilon_{\mu \nu \alpha \beta} 
\bar \ell \sigma^{\mu\nu}\ell \, \Big[ G \varepsilon^{\ast\alpha} k^\beta 
+ H \varepsilon^{\ast\alpha} q^\beta + 
N (\varepsilon^\ast q) q^\alpha k^\beta \Big] \\
&+& i \,\bar \ell \sigma_{\mu\nu}\ell \, \Big[
G_1 (\varepsilon^{\ast\mu} k^\nu - \varepsilon^{\ast\nu} k^\mu) + 
H_1 (\varepsilon^{\ast\mu} q^\nu - \varepsilon^{\ast\nu} q^\mu) +
N_1 (\varepsilon^\ast q) (q^\mu k^\nu - q^\nu k^\mu) \Big] \Bigg\}~,\nnb
\eea
where
\bea
A_1 &=& \frac{1}{q^2} \Big( C_{BR} + C_{SL} \Big) g_1 +
\Big( C_{LL}^{tot} + C_{RL} \Big) g ~, \nnb \\
A_2 &=& \frac{1}{q^2} \Big( C_{BR} - C_{SL} \Big) f_1 +
\Big( C_{LL}^{tot} - C_{RL} \Big) f ~, \nnb \\
B_1 &=& \frac{1}{q^2} \Big( C_{BR} + C_{SL} \Big) g_1 +
\Big( C_{LR}^{tot} + C_{RR} \Big) g ~, \nnb \\
B_2 &=& \frac{1}{q^2} \Big( C_{BR} - C_{SL} \Big) f_1 +
\Big( C_{LR}^{tot} - C_{RR} \Big) f ~, \nnb \\
G &=& 4 C_T g_1 ~, \nnb \\
N &=& - 4 C_T \frac{1}{q^2} (f_1+g_1) ~, \\
H &=& N (qk) ~, \nnb \\
G_1 &=& - 8 C_{TE} g_1 ~, \nnb \\
N_1 &=& 8 C_{TE} \frac{1}{q^2} (f_1+g_1) ~, \nnb \\ 
H_1 &=& N_1(qk) \nnb
\eea
For the inner Bremsstrahlung part we get
\bea
\label{ib}
{\cal M}_{IB} &=& \frac{\alpha G_F}{4 \sqrt{2} \, \pi} V_{tb} V_{tq}^*  
e f_B i \,\Bigg\{
F\, \bar \ell  \Bigg(
\frac{{\not\!\varepsilon}^\ast {\not\!p}_B}{2 p_1 k} - 
\frac{{\not\!p}_B {\not\!\varepsilon}^\ast}{2 p_2 k} \Bigg) 
\gamma_5 \ell \nnb \\
&+& F_1 \, \bar \ell  \Bigg[
\frac{{\not\!\varepsilon}^\ast {\not\!p}_B}{2 p_1 k} -
\frac{{\not\!p}_B {\not\!\varepsilon}^\ast}{2 p_2 k} +
2 m_\ell \Bigg(\frac{1}{2 p_1 k} + \frac{1}{2 p_2 k}\Bigg)
{\not\!\varepsilon}^\ast \Bigg] \ell \Bigg\}~.
\eea
In obtaining this expression we have used
\bea
\la 0 \ve \bar s \gamma_\mu \gamma_5 b \ve B \ra &=& 
-~i f_B p_{B\mu}~, \nnb \\
\la 0 \ve \bar s \sigma_{\mu\nu} (1+\gamma_5) b \ve B \ra &=& 0~,\nnb
\eea
and conservation of the vector current.
The functions $F$ and $F_1$ are defined as follows
\bea
F &=& 2 m_\ell \Big( C_{LR}^{tot} - C_{LL}^{tot} + C_{RL} - C_{RR} \Big)
+ \frac{m_B^2}{m_b}
\Big( C_{LRLR} - C_{RLLR} - C_{LRRL} + C_{RLRL} \Big)~, \nnb \\
F_1 &=&\frac{m_B^2}{m_b} \Big( C_{LRLR} - C_{RLLR} + C_{LRRL} - C_{RLRL}
\Big)~,
\eea
The double differential decay width of the $B \rar \ell^+ \ell^- \gamma$
process in the rest frame of the $B$ meson is found to be

\bea
\frac{d \Gamma}{dE_\gamma\, dE_1} = \frac{1}{256 \pi^3 m_B}
\vel {\cal M} \ver^2~,
\eea 
where $E_\gamma$ and $E_1$ are the photon and one of the final lepton energy,
respectively.
The boundaries of $E_\gamma$ and $E_1$ are determined from the following
inequalities
\bea
0 \leq E_\gamma \leq \frac{m_B^2 - 4 m_\ell^2}{2 m_B}~,\nnb
\eea
\bea
\frac{m_B - E_\gamma}{2} - \frac{E_\gamma}{2} v \leq E_1 \leq
\frac{m_B - E_\gamma}{2} + \frac{E_\gamma}{2} v ~,
\eea
where 
\bea
v=\sqrt{1 - \frac{4 m_\ell^2}{q^2}}~,\nnb 
\eea
is the lepton velocity.

The $\vel {\cal M}_{SD} \ver^2$ term is infrared free; interference term has
an integrable infrared singularity and only $\vel {\cal M}_{IB} \ver^2$ term
has infrared singularity due to the emission of soft photon.
In the soft photon limit the $B_q \rar \ell^+ \ell^- \gamma$ decay cannot be
distinguished from the pure leptonic $B_q \rar \ell^+ \ell^-$ decay. For
this reason, in order to obtain a finite result the $B \rar \ell^+ \ell^-
\gamma$ and the pure leptonic $B_q \rar \ell^+ \ell^-$ decay with radiative
corrections must be considered together. It was shown explicitly in the second
reference of \cite{R10} that when both processes are considered together,
all infrared singularities coming from the real photon emission and the
virtual photon corrections are indeed canceled and the final result is
finite. In the present work our point of view is slightly different from the
standard description, namely, we consider the $B_q \rar \ell^+ \ell^-
\gamma$ decay as a different process but not as the ${\cal O}(\alpha)$
correction to the $B \rar \ell^+ \ell^-$ decay. In other words, we consider
the photon in the $B_q \rar \ell^+ \ell^- \gamma$ decay as a hard photon.
For this reason, in order to obtain the decay width of the $B_q \rar \ell^+
\ell^- + (hard~photon)$ we must impose a cut on the photon energy, which
will correspond to the experimental cut imposed on the minimum energy for
detectable photon. We require the photon energy to be larger than $25~MeV$,
i.e., $E_\gamma \ge \delta m_B/2$, where $\delta \ge 0.010~GeV$.      

After integrating over lepton energy, we get the following expression for
the photon energy distribution
\bea
\label{bela}
\lefteqn{
\frac{d\Gamma}{dx} = - \,\vel \frac{\alpha G_F}{4 \sqrt{2} \, \pi} V_{tb} V_{tq}^* \ver^2 \,
\frac{\alpha}{\ga 2 \, \pi \dr^3}\,\frac{\pi}{4}\,m_B
\,\Bigg( x^3 v\, \Bigg\{
4 m_\ell \, 
\mbox{\rm Re}\Big( [A_1+B_1] G^\ast\Big)
- \, 4 m_B^2 r \, \mbox{\rm Re}\Big( A_1 B_1^\ast + A_2 B_2^\ast \Big)} \nnb \\
&&- \, 4 \,\Big[ \vel H_1 \ver^2 (1-x) + \mbox{\rm Re}\Big( G_1 H_1^\ast \Big) x
\Big] \frac{(1+ 8 r - x)}{x^2} \nnb \\
&&- \, 4 \, \Big[ \vel H \ver^2 (1-x) + \mbox{\rm Re}\Big( G H^\ast \Big) x
\Big] \frac{(1- 4 r - x)}{x^2} \nnb \\
&&+ \, \frac{1}{3} m_B^2 \Big[ 2\,\mbox{\rm Re}\Big( G N^\ast \Big) + 
m_B^2 \vel N \ver^2 (1-x) \Big] (1-4r-x)  \\
&&+ \, \frac{1}{3} m_B^2 \Big[ 2\,\mbox{\rm Re}\Big( G_1 N_1^\ast \Big) + 
m_B^2 \vel N_1 \ver^2 (1-x) \Big] (1+8r-x) \nnb \\
&&- \, \frac{2}{3} m_B^2 \Big( \vel A_1 \ver^2 + \vel A_2 \ver^2 +
\vel B_1 \ver^2 + \vel B_2 \ver^2\Big) 
( 1 - r - x) - \frac{4}{3} \Big( \vel G \ver^2 + \vel G_1 \ver^2 \Big)
\frac{(1 + 2 r - x)}{(1 - x)}\nnb \\
&&+\,2 m_\ell \,\mbox{\rm Im}\Big([A_2+B_2] [6 H_1^\ast (1-x) 
 + 2 G_1^\ast x - m_B^2 \,N_1^\ast  x (1-x)]\Big)\frac{1}{x} \Bigg\} \nnb \\
&&+ \, 4 f_B \,\Bigg\{ 2 v \,\Bigg[
\mbox{\rm Re}\Big(F G^\ast\Big)\frac{1}{(1-x)}
- \mbox{\rm Re}\Big(F H^\ast\Big)
+ \,m_B^2\, \mbox{\rm Re}\Big(F N^\ast\Big)
+ m_\ell \,\mbox{\rm Re}\Big([A_2+B_2] F_1^\ast\Big)
\Bigg] \, x (1-x)\nnb \\
&&+\, {\rm ln} \frac{1 + v}{1 - v} \Bigg[
m_\ell \, \mbox{\rm Re}\Big([A_2+B_2] F_1^\ast\Big) \, x (x-4 r)
+ 2  \, \mbox{\rm Re}\Big(F H^\ast \Big) 
\Big[1 -x + 2 r (x-2) \Big] \nnb \\
&&- \, 4 r x\, \mbox{\rm Re}\Big(F G^\ast\Big)  
+ m_B^2 \, \mbox{\rm Re}\Big(F N^\ast\Big) \, x (x-1)
- m_\ell \,\mbox{\rm Re}\Big([A_1+B_1] F^\ast\Big) \, x^2 
\Bigg]\Bigg\} \nnb \\
&&+ \,  4 f_B^2 
\Bigg\{2 v\, \Big( \vel F \ver^2+ 
(1-4 r) \vel F_1 \ver^2 \Big) \frac{(1-x)}{x} \nnb \\
&&+ \, {\rm ln}\frac{1 + v}{1 - v} \Bigg[ \vel F \ver^2 \Big( 2 +\frac{4 r}{x} -\frac{2}{x} -x \Big)
+ \vel F_1 \ver^2 \Bigg( 2 (1-4 r) - \frac{2 \ga 1- 6 r + 8 r^2 \dr}{x} -x
\Bigg) \Bigg]\Bigg\}\Bigg)~,\nnb
\eea
where $x=2 E_\gamma/m_B$ is the dimensionless photon energy,
$r=m_\ell^2/m_B^2$.

It follows from Eq. (\ref{bela}) that in order to calculate the decay width
explicit forms of the form factors $g,~f,~g_1$ and $f_1$ are needed. 
These form factors are calculated in framework of light--cone $QCD$ sum rules 
in \cite{R4} and \cite{R10}, and their $q^2$ dependences, to a very good accuracy,
can be represented in the following dipole forms,
\bea
\label{ff}
g_(q^2) &=& \frac{1~GeV}{(1-\displaystyle{\frac{q^2}{5.6^2})^2}}~,
~~~~~~~~~~
f_(q^2) = \frac{0.8~GeV}{(1-\displaystyle{\frac{q^2}{6.5^2})^2}}~,
 \nnb \\ \nnb  \\ \nnb \\
g_1(q^2) &=& \frac{3.74~GeV^2}{(1-\displaystyle{\frac{q^2}{40.5})^2}}~,
~~~~~~~~~
f_1(q^2) = \frac{0.68~GeV^2}{(1-\displaystyle{\frac{q^2}{30})^2}}~,
\eea
which we will use in the numerical analysis.

\section{Numerical analysis and discussion}
In this section we will present our numerical analysis. Numerical results
are presented only for the $B_s \rar \ell^+ \ell^- \gamma$ decay. It is
clear that in the SU(3) limit the difference between the decay rates is attributed to
the CKM matrix elements only, i.e.,
\bea
\frac{\Gamma(B_d \rar \ell^+ \ell^- \gamma)}
{\Gamma(B_s \rar \ell^+ \ell^- \gamma)} \simeq \vel \frac{V_{tb} V_{td}^\ast}
{V_{tb} V_{ts}^\ast} \ver^2 \simeq \frac{1}{20}~.\nnb
\eea 
The values of the main input parameters which have been used in the present work are:
$m_b=4.8~GeV$, $m_c=1.35~GeV$, $m_\tau=1.78~GeV$, 
$\vel V_{tb} V_{ts}^\ast \ver = 0.045$, $\alpha^{-1}=137$, 
$G_F=1.17\times 10^{-5}~GeV^{-2}$. For the Wilson coefficients
$C_7^{eff}(m_b)$ and $C_{10}(m_b)$ we have used the results given in 
\cite{R12,R13}. In the leading logarithmic approximation, at the scale
${\cal O}(\mu=m_b)$ they are given as
$C_7^{eff}(m_b)=-0.315$, $C_{10}(m_b)=4.6242$. Although individual Wilson
coefficients at $\mu \sim m_b$ level are all real,
the effective Wilson coefficient
$C_{9}^{eff}(m_{b})$ has a finite phase. The analytic expression of  
$C_{9}^{eff}(m_{b})$ for the $b\rar s$ transition, in next--to--leading order
approximation is given as
\bea
\label{c9}
C_9^{eff}(m_b,\hat s) &=& C_9(m_b) + 0.124 w(\hat s) + g(\hat m_c, \hat s) 
(3 C_1 + C_2 + 3 C_3 + C_4 + 3 C_5 + C_6) \nnb \\
&-& \frac{1}{2} g(\hat m_q, \hat s)
(C_3 + 3 C_4) - \frac{1}{2} g(\hat m_b, \hat s) (4 C_3 + 4 C_4 + 3 C_5 +
C_6)\nnb \\
&+& \frac{2}{9} (3 C_3 + C_4 + 3 C_5 + C_6)~,
\eea 
where $m_q=m_q/m_b$, $\hat s = q^2/m_b^2$ and the values of the individual Wilson
coefficients are listed in Table 1.
\begin{table}[ht]  
\renewcommand{\arraystretch}{1.5}
\addtolength{\arraycolsep}{3pt}
$$
\begin{array}{|c|c|c|c|c|c|c|c|c|}
\hline
C_{1} & C_{2} & C_{3} & C_{4} & C_{5} & C_{6}
& C_{7}^{eff} &    
C_{9} & C_{10}^{eff}  \\ \hline
-0.248 & 1.107 & 0.011 & -0.026 & 0.007 & -0.031 & -0.315 & 4.344 & -4.6242
\\ \hline 
\end{array}
$$
\caption{The numerical values of the Wilson coefficients at $\mu\sim m_{b}$
scale within the SM.}
\renewcommand{\arraystretch}{1}
\addtolength{\arraycolsep}{-3pt}
\end{table}
In Eq. (\ref{c9}) $w(\hat s)$ describes one gluon corrections to the matrix
element of the operator ${\cal O}_9$ and the function 
$g(\hat m_q, \hat s)$ stands for the one loop corrections to the four quark
operators ${\cal O}_1$--${\cal O}_6$ with mass $m_{q}$ at the dilepton
invariant mass $s$ \cite{R14,R15}:
\bea
\lefteqn{
g \ga \hat m_q,\hat s^\prime \dr = - \frac{8}{9} \ln \hat m_q +
\frac{8}{27} + \frac{4}{9} y_q -
\frac{2}{9} \ga 2 + y_q \dr \sqrt{\vel 1 - y_q \ver}} \nnb \\
&&\times \Bigg[ \Theta(1 - y_q)
\ga \ln \frac{1  + \sqrt{1 - y_q}}{1  -  \sqrt{1 - y_q}} - i \pi \dr
+ \Theta(y_q - 1) \, 2 \, \arctan \frac{1}{\sqrt{y_q - 1}} \Bigg], \nnb
\eea
where  $y_q=4 \hat m_q^2/\hat s^\prime$ and $\hat s^\prime = q^2/m_b^2$.  
It is well known that the Wilson coefficient $C_9^{eff}$ receives also long distance
contributions,
which have their origin in the real $c\bar c$
intermediate states, i.e., $J/\psi$, $\psi^\prime$,
$\cdots$ (see \cite{R16}). In this work we restrict ourselves only to
short contributions. Furthermore we assume that 
all new Wilson coefficients are real and varied in the region $-4 \le C_X \le +4$.   

In Fig. (1) we present the dependence of the integrated branching ratio of
the $B \rar \tau^+ \tau^- \gamma$ decay on the new Wilson coefficients for
the cut $\delta=0.01$ imposed on the photon energy, without long
distance effects. It clearly follows from this figure that as the new
Wilson coefficients $C_T,~C_{RL},~C_{LR},~C_{LRLR}$ and $C_{RLRL}$ increase
from $-4$ to $+4$ branching ratio decreases. However this behavior is
reversed for the coefficients $C_{LL},~C_{RR},~C_{LRRL}$ and
$C_{RLLR}$ i.e., when these
coefficients increase from $-4$ to $+4$ branching ratio also increases
accordingly. Exception to these cases takes place for the coefficient
$C_{TE}$. In the region $-4\le C_{TE} \le 0$ branching ratio decreases 
and in the region $0\le C_{TE} \le +4$ it tends to increase.
 
For the choice of the photon energy cut $\delta=0.02$ all the previous
arguments remain valid with only a slight decrease in the value of the
branching ratio.

From all present figures we observe that when all Wilson coefficients lie in
the range $-4 \le C_X \le -2$, the branching ratio is more sensitive to the
existence of tensor $C_T$, scalar $C_{LRLR},~C_{RLRL}$ and vector $C_{LL}$ 
type interactions. On the other side, when Wilson coefficients lie in the
region $+2 \le C_X \le +4$ the branching ratio is more sensitive to the
scalar type interaction with coefficients $C_{LRRL}$ and $C_{RLLR}$.

Photon energy distribution can also give useful information about new
physics effects. For this purpose, in Fig. (2) we present the dependence of 
the differential branching ratio for the $B \rar \tau^+ \tau^- \gamma$ decay 
on the dimensionless variable
$x=2 E_\gamma/m_B$  at different values of tensor interaction with coefficient        
$C_T$. We observe from this figure that when $C_T < 0$ then the related
tensor interaction gives constructive contribution to the SM result, and
when $C_T > 0$ the contribution is destructive. In other words measurement
of the differential branching ratio can give essential information about the
sign of new Wilson coefficients. 

Performing measurement at different photon energies can give
information not only about magnitude but also about the sign of the
new Wilson coefficient interaction. 

Note that the results presented in this work can easily be applied to the
$B_s \rar \mu^+ \mu^- \gamma$ decay. For example, the branching
ratio for the $B_s \rar \mu^+ \mu^- \gamma$ decay at $\delta=0.01$, without
the long distance effects at $C_{TE} = C_T = \pm 4$ is larger
about 5 times, compared to that of the SM prediction of the branching ratio
for the $B_s \rar \mu^+ \mu^- \gamma$ decay. 
Additionally, the dependence of the branching ratio on the 
new Wilson coefficients is symmetric with respect to the zero point (see
Fig. (3)). It should be stressed that by studying the Dalitz distribution 
$d \Gamma/dE_\gamma dE_1$ at different fixed values of the final lepton (or
photon) energies we can get useful information not only about the magnitude
of the new Wilson coefficients but also about their sign.     

In conclusion, using a general, model independent effective Hamiltonian, the
$B_s \rar \ell^+ \ell^- \gamma$ decay is studied. It has been shown that the
branching ratio and photon energy distribution are very sensitive to the
existence of new physics beyond SM. We conclude that the radiative $B_s \rar
\ell^+ \ell^- \gamma$ decay can be measured ib the B factories as well as
LHC--B experiments, in which $\approx 2 \times 10^{11}$ $B_s$ mesons are
expected to be produced per year.

\newpage

\section*{Figure captions}
{\bf Fig. (1)} The dependence of the integrated branching ratio of
the $B_s \rar \tau^+ \tau^- \gamma$ decay on the new Wilson coefficients for
the cut $\delta=0.01$ imposed on the photon energy, only for short distance 
effects.\\ \\
{\bf Fig. (2)} The dependence of the differential branching ratio for the 
$B_s \rar \tau^+ \tau^- \gamma$ decay on the dimensionless variable
$x=2 E_\gamma/m_B$ at different values 
of tensor interaction with coefficient $C_T$, without the long distance
effects.\\ \\
{\bf Fig. (3)} The same as in Fig. (1), but for the $B_s \rar \mu^+ \mu^-
\gamma$ decay.

\newpage

\begin{figure}
\vskip 1cm
    \includegraphics{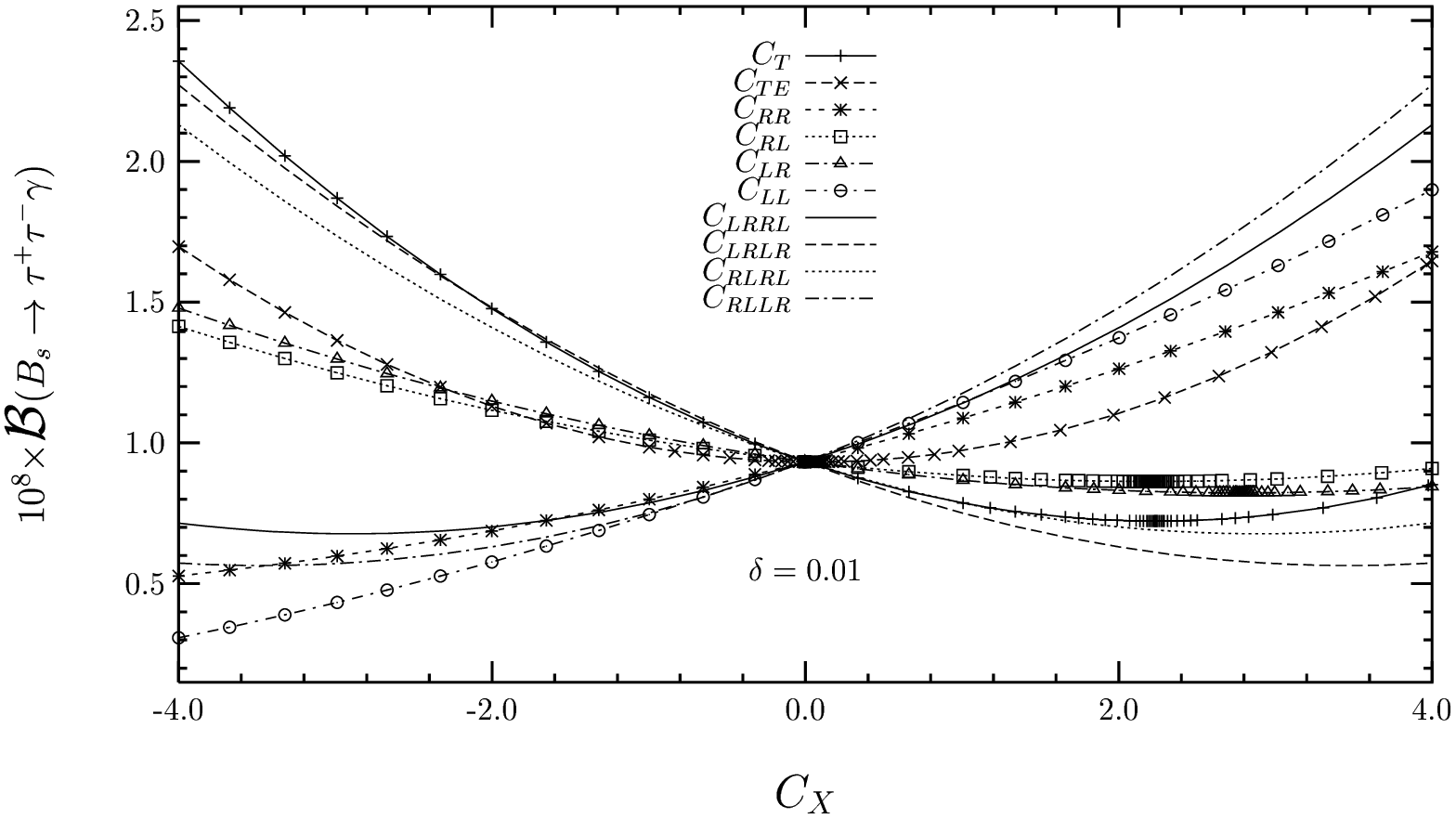}
\vskip 8.1cm
\caption{}
\end{figure}

\begin{figure}  
\vskip 1.5 cm
    \includegraphics{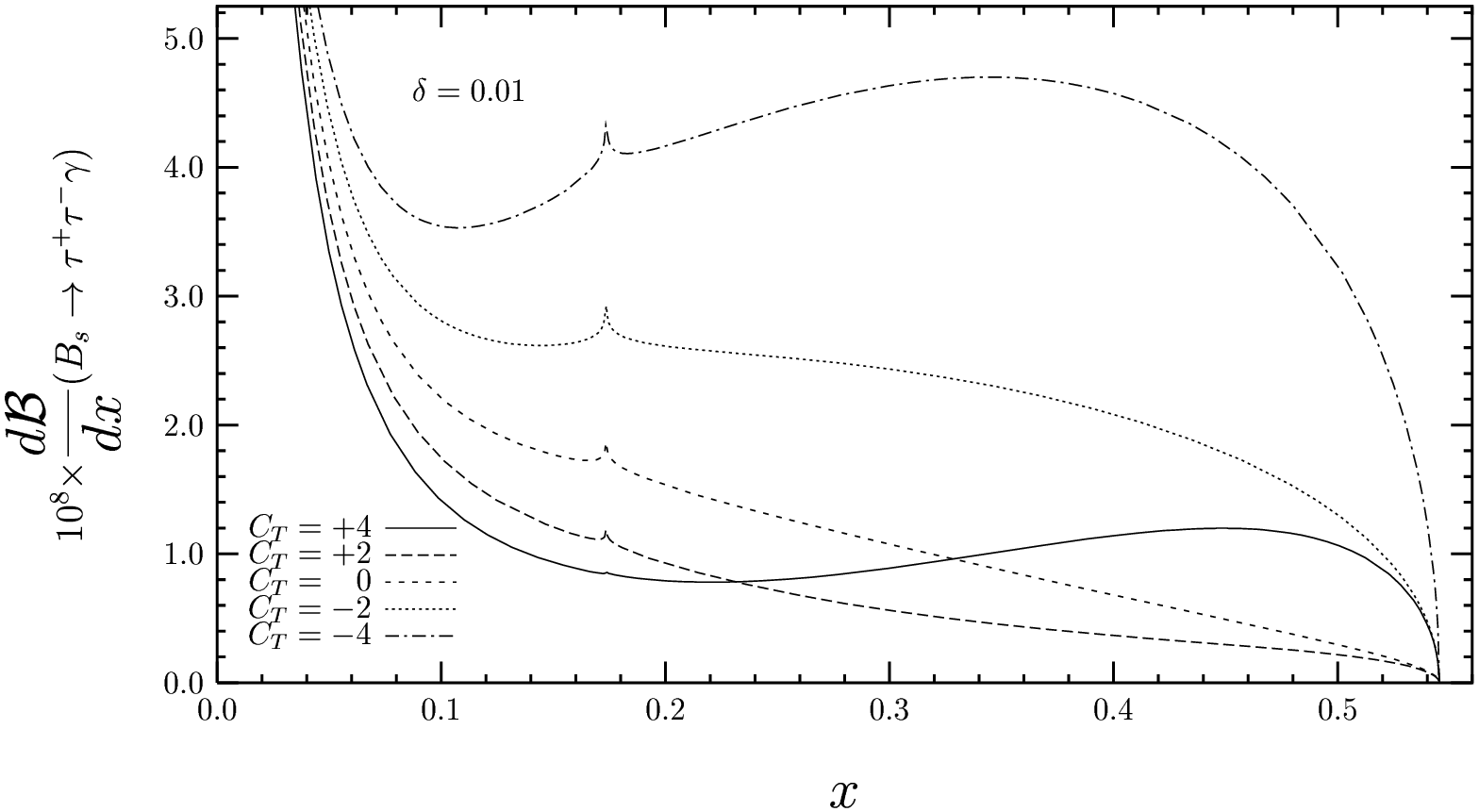}
\vskip 9. cm   
\caption{}
\end{figure}

\begin{figure}
\vskip 1.5 cm
    \includegraphics{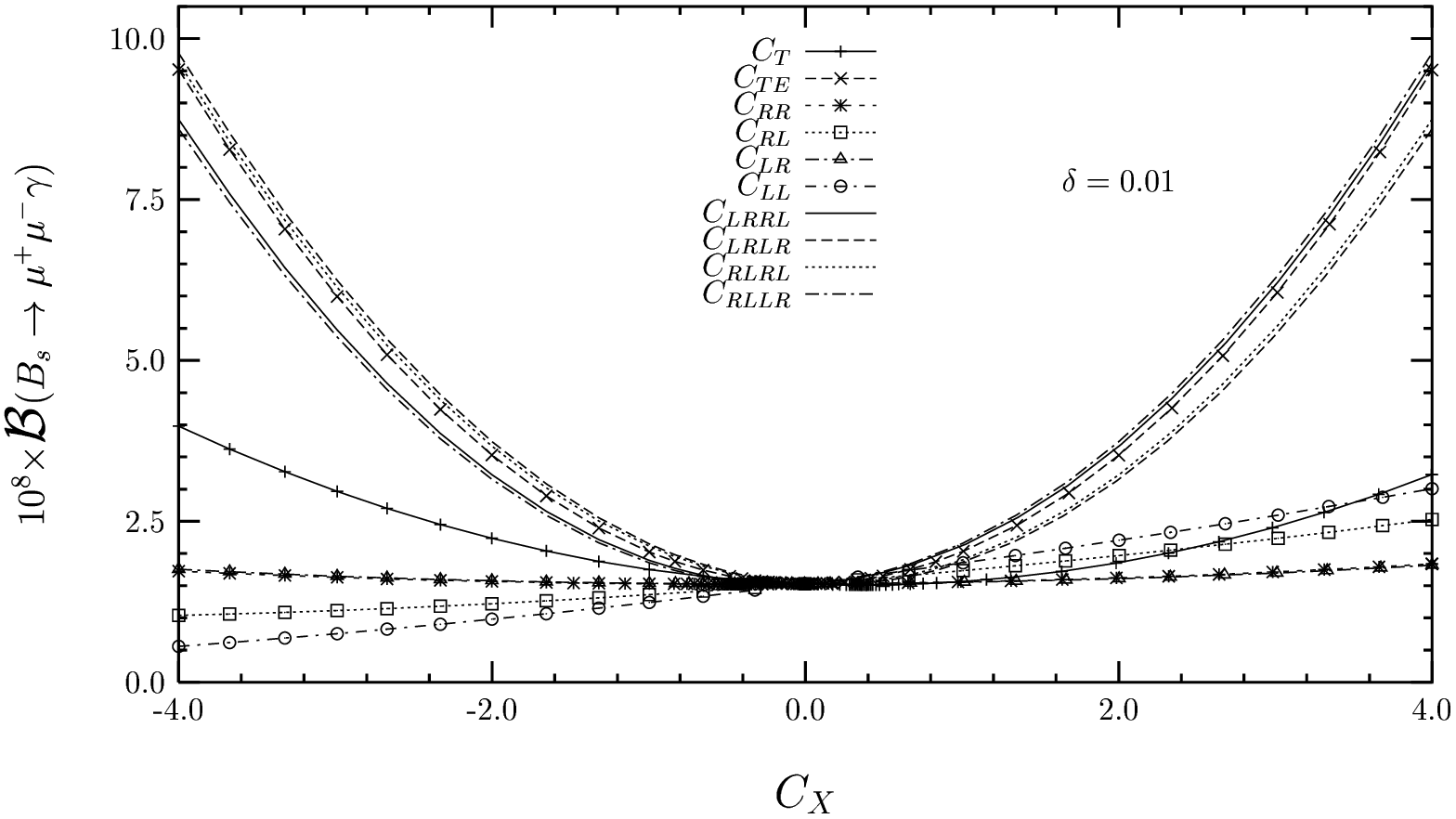}
\vskip 9. cm
\caption{}
\end{figure}

\end{document}